\documentclass[twocolumn]{article}
\usepackage{natbib}
\usepackage[dvips]{graphicx}
\usepackage{amsmath}
\usepackage{amsfonts}
\usepackage{amsbsy}
\usepackage{amssymb}

\def\bfgr#1{\boldsymbol{#1}}
\def\cP{{\mathcal{P}}}
\def\cH{{\mathcal{H}}}
\def\br{\bar{r}}
\def\lb#1{\label{#1}}
\def\cbP{{\bfgr{\mathcal{P}}}}
\def\p{\partial}

\def\dfrac#1#2{{\displaystyle\frac{#1}{#2}}}
\def\be{\begin{equation}}
\def\ee{\end{equation}}
\def\ba#1{\begin{array}{#1}}
\def\ea{\end{array}}
\def\bea{\begin{eqnarray}}
\def\eea{\end{eqnarray}}
\def\df{{\mathrm{d}}}
\def\I{{\mathcal{I}}}
\def\J{{\mathcal{J}}}
\def\div{{\mathrm{div}}}
\def\curl{{\mathrm{curl}}}

\def\E{{\mathbf{E}}}
\def\B{{\mathbf{B}}}
\def\D{{\mathbf{D}}}
\def\H{{\mathbf{H}}}
\def\L{{\mathcal{L}}}

\def\combss#1#2#3{{\mathsurround 0pt\hbox to 0pt {\hspace*{#3}\raisebox{#2}{${\scriptscriptstyle #1}$}\hss}}}
\def\Ce{\combss{e}{0.45ex}{0.2em}C}
\def\Cm{\combss{m}{0.45ex}{0.1em}C}

\begin{document}

\author{\textbf{Alexander A. Chernitskii}}

%\title{NL3572---Born--Infeld equations}
\title{\textbf{BORN--INFELD EQUATIONS\footnote%
{Entry in Encyclopedia of Nonlinear Science, ed. Alwyn Scott. New York and London: Routledge, 2004, pp. 67-69;
http://www.routledge-ny.com/ref/nonlinearsci/ .}}}
\date{}

%\IDnumber{NL3572}
\maketitle

Classical linear vacuum electrodynamics with point massive charged
particles has two limiting properties: the electromagnetic energy of a
point particle field is infinity, and a Lorentz force must be postulated
to describe interactions between point particles and an electromagnetic
field.  Nonlinear vacuum electrodynamics can be free of these
imperfections.

Gustav \citet{NL3572Mie}
considered a nonlinear electrodynamics model in the framework of
his ``Fundamental unified theory of matter''. In this theory the electron
is represented by a nonsingular solution with a finite electromagnetic
energy, but Mie's field equations are non-invariant under the gauge
transformation for an electromagnetic four-potential
(addition of the four-gradient of an arbitrary scalar function).

\begin{figure}
\begin{center}
\includegraphics[width=0.8\linewidth]{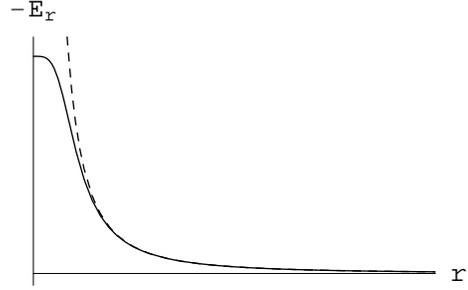}
\end{center}
\caption{Radial components of electrical field
for Born's electron and purely Coulomb field (dashed).
\label{BElectronPot}}
\end{figure}

Max \citet{NL3572Born1934} considered
a nonlinear electrodynamics model that is invariant under the gauge
transformation. A stationary electron in this model is represented by an
electrostatic field configuration that is everywhere finite, in contrast
to the case of linear electrodynamics when the electron's field is
infinite at the singular point (see Figure \ref{BElectronPot}). The
central point in Born's electron is also singular because there is a
discontinuity of electrical displacement field at this point (hedgehog
singularity). The full electromagnetic energy of this electron's field
configuration is finite.

Born and Leopold Infeld (1934) then considered
a more general nonlinear electrodynamics model which has the same solution
associated with electron. Called
Born--Infeld electrodynamics, this model is based on Born--Infeld
equations, which have the form of  Maxwell's equations, including
electrical and magnetic field strengths $\E$, $\H$, and inductions $\D$,
$\B$ with nonlinear constitutive relations
$\D=\D(\E,\B)$, $\H=\H(\E,\B)$ of a special kind.
For inertial reference frames and in the region outside of field
singularities, these equations are
\begin{equation}
\left\{
\begin{array}{rl}
\div \B  = 0\;,&\\[2.5mm]
\dfrac{1}{c}\,\dfrac{\p\B}{\p t}  {}+{}  \curl\E  = 0\;,&\\[2.5mm]
\div \D  = 0\;,&\\[2.5mm]
\dfrac{1}{c}\,\dfrac{\p\D}{\p t}  {}-{}  \curl\H  = 0\;,&
\end{array}
\right.
%\quad,
\label{Eq:Maxwell}
\end{equation}
where
\begin{align}
%\begin{eqnarray}
\nonumber
&\D = \dfrac{1}{\L}\,(\E  {}+{}  \chi^2\,\J \B)
\quad,
\\
%\qquad
&\H = \dfrac{1}{\L}\,(\B  {}-{}  \chi^2\,\J \E)
\label{MatEq}
\quad,
\end{align}
\begin{align}
\nonumber
&\L {}\equiv{}   \sqrt{|\,1 {}-{}  \chi^2\,\I  {}-{}  \chi^4\,\J^2\,|}
\;,
\\
%\quad
&\I = \E\cdot\E  {}-{}  \B\cdot\B
\;,\quad
\J = \E\cdot\B
\;.
\label{Def:Lagrangian}
\end{align}
%\end{eqnarray}
%
Relations (\ref{MatEq}) can be resolved for $\E$ and $\H$:
\begin{align}
\nonumber
&\E =   \dfrac{1}{\cH}\left(
\D {}-{} \chi^2\,\cbP {}\times{} \B
\right)
\quad,
\\
%\qquad
&\H =  \dfrac{1}{\cH}\left(
\B {}+{} \chi^2\,\cbP {}\times{} \D
\right)
\quad,
\lb{Expr:EHDB}
\end{align}
where
\mbox{$\cH {}={} \sqrt{1 {}+{} \chi^2 \left(\D^2 {}+{} \B^2 \right)
{}+{} \chi^4\,\cbP^2}$} ,
\mbox{$\cbP \equiv  \D {}\times{} \B$} .
Using relations (\ref{Expr:EHDB}) for equations (\ref{Eq:Maxwell}),
the fields $\D$ and $\B$ are unknown.

The symmetrical energy-momentum tensor for Born--Infeld equations have the
following components:
\begin{align}
\nonumber
&T^{00} = \frac{1}{4\pi\,\chi^2}\left(\cH {}-{} 1 \right)
\;,\quad
T^{0i} = \frac{1}{4\pi}\,\cP^i
\;,
\\
\nonumber
&T^{ij} = \frac{1}{4\pi}\left\{\delta^{ij}
\left[\D\cdot\E {}+{} \B\cdot\H {}-{}
\chi^{-2} \left(\cH {}-{} 1\right)
\right]\right.
\\
&\quad\quad\quad{}-{}
\left.\left(D^i\,E^j {}+{} B^i\,H^j\right)\right\}
\;.\quad
\label{Def:TEM}
\end{align}

In spherical coordinates the field of static Born's electron solution may
have only radial components
\begin{equation}
D_r  {}={}  \frac{e}{r^2}\;,\quad
E_r  {}={}  \frac{e}{\sqrt{\br^4 {}+{} r^4}}\;,\quad
\label{Sol:Stat}
\end{equation}
where $e$ is the electron's charge and $\br {}\equiv{} \sqrt{|\chi\,e|}$.
At the point $r=0$ the electrical field has the maximum absolute value
\begin{equation}
\label{MaxElField}
|E_r(0)|=\frac{|e|}{\br^2}=\frac{1}{\chi} \,,
\end{equation}
which Born and Infeld called the absolute field constant.

The energy of field configuration (\ref{Sol:Stat}) is
\begin{equation}
\label{BornElectronEnergy}
m = \int T^{00}\,{\rm d}V
= \frac{2}{3}\,\beta\,\frac{\br^3}{\chi^2}
\quad,
\end{equation}
where the volume integral
is calculated over the whole space, and
\begin{equation}
\beta {}\equiv{} \int\limits_0^\infty\frac{\df r}{\sqrt{1 {}+{} r^4}} {}={}
\frac{\left[\Gamma (\frac{1}{4})\right]^2}{4\,\sqrt{\pi}}
\approx 1.8541
\;.
\end{equation}
In view of the definition for $\br$ below (\ref{Sol:Stat}),
Equation (\ref{BornElectronEnergy}) yields
\begin{equation}
\label{BornElectronRadius}
\br = \frac{2}{3}\,\beta\,\frac{e^2}{m}
\quad.
\end{equation}
Considering  $m$ as the mass of electron and using (\ref{MaxElField}),
\citet{NL3572BornInfeld1934a}
estimated the absolute field constant $\chi^{-1}\approx 3 \cdot 10^{20}\,\mbox{V}/\mbox{m}$.
Later \citet{NL3572BornSchroding1935} gave a new estimate
(two orders of magnitude less)
based on some considerations taking into account the spin of electron.
(Of course, such estimates may be corrected with more detailed models.)

An electrically charged solution of the Born--Infeld equations can be
generalized to a solution with the singularity having both electrical
and magnetic charges \citep{NL3572Chernitskii1999}. A corresponding
hypothetical particle is called a dyon \citep{NL3572Schwinger1969}.
Nonzero (radial) components of fields for this solution have the form
\begin{align}
\nonumber
D_r  {}={}  \frac{\Ce}{r^2}\;,\quad
E_r  {}={}  \frac{\Ce}{\sqrt{\br^4 {}+{} r^4}}\;,\quad
\\
B_r  {}={}  \frac{\Cm}{r^2}\;,\quad
H_r  {}={}  \frac{\Cm}{\sqrt{\br^4 {}+{} r^4}}\;,\quad
\label{Sol:StatDyon}
\end{align}
where $\Ce$ is electric charge and $\Cm$ is magnetic one,
$\br {}\equiv{}
\left[\chi^2\left(\Ce^2 {}+{} \Cm^2\right)\right]^{{1}/{4}}$.
The energy of this solution is given by formula (\ref{BornElectronEnergy})
with this definition for $\br$.
It should be noted that space components of electromagnetic potential for the static dyon
solution have a line singularity.

A generalized Lorentz force appears when a small
almost constant field $\tilde{\D},\,\tilde{\B}$ is considered
in addition to the moving dyon solution.
The sum of the field $\tilde{\D},\,\tilde{\B}$ and the field of the dyon
with varying velocity is taken as an initial approximation to some exact solution.
Conservation of total momentum gives the following
trajectory equation \citep{NL3572Chernitskii1999}:
\begin{multline}
\lb{Eq:trajectfirst}
m\,\frac{\df }{\df t}\,\frac{\bf v}{\sqrt{1 {}-{} {\bf v}^2}}
=
\Ce \left(\tilde{\D} {}+{} {\bf v} {}\times{} \tilde{\B}
\right)
\\
 {}+{}
\Cm \left( \tilde{\B} {}-{} {\bf v} {}\times{} \tilde{\D}
\right)
\quad,
\end{multline}
where ${\bf v}$ is the velocity of the dyon,
$m$ is the energy for static dyon defined by (\ref{BornElectronEnergy}).

A solution with two dyon singularities (called a bidyon) having equal
electric ($\Ce=e/2$) and opposite magnetic charges can be considered as a
model for a charged particle with spin \citep{NL3572Chernitskii1999}. Such
a solution has both angular momentum and magnetic moment.

A plane electromagnetic wave with arbitrary
polarization and form in the direction of propagation (without coordinate
dependence in a perpendicular plane) is an exact solution to Born--Infeld
equations. The simplest case assumes one nonzero component of the vector
potential ($A_y\equiv\phi(t,x)$), whereupon Equations (1) reduce to the
linearly polarized plane wave equation
\begin{multline}
\label{BIeq}
\left(1+ \chi^2\,\phi_x^2\right)\phi_{tt} - \chi^2\,2\,\phi_x\,\phi_t\,\phi_{xt}
\\
- \left(c^2- \chi^2\,\phi_t^2\right)\phi_{xx}
=0
\end{multline}
with indices indicating partial derivatives.
Sometimes called the Born--Infeld equation, Equation (\ref{BIeq})
has solutions $\phi=\zeta(x^1 -x^0)$ and $\phi=\zeta(x^1 + x^0)$, where
$\zeta(x)$ is arbitrary function \citep{NL3572Whitham1974} . Solutions
comprising two interacting waves propagating in opposite directions is
obtained via a hodograph transform \citep{NL3572Whitham1974}.
\citet{NL3572BrunelliAshok1998} have found a Lax representation for
solutions of this equation.

A solution to the Born--Infeld equations which is the sum of two
circularly polarized waves propagating in different directions was
obtained by Erwin \citet{NL3572Schrodinger1943b}.

Equations (\ref{Eq:Maxwell}) with relations (\ref{MatEq})
have an interesting characteristic equation
\citep{NL3572Chernitskii1998b}:
\begin{equation}
\label{CharEq}
\mathfrak{g}^{\mu\nu}\,\frac{\p \Phi}{\p x^\mu}\,\frac{\p \Phi}{\p x^\nu}=0
\;,\quad
\mathfrak{g}^{\mu\nu} \equiv g^{\mu\nu} - 4\pi\,\chi^2\,T^{\mu\nu}
\;,
\end{equation}
where $\Phi (x^\mu)=0$ is an equation of the characteristic surface and
$T^{\mu\nu}$ are defined by (\ref{Def:TEM}). This form for $\mathfrak{g}^{\mu\nu}$,
including additively the energy-momentum tensor, is special for
Born--Infeld equations.

The Born--Infeld model appears also
in quantized string theory \citep{NL3572FradkinTseytlin1985} and in
Einstein's unified field theory with a nonsymmetrical metric
\citep{NL3572ChernikovChavokhina}. In general, this nonlinear
electrodynamics model is connected with ideas of  space-time
geometrization and general relativity
\citep[see][]{NL3572Eddington,NL3572Chernitskii2002}.

%\makesignature

\vspace{.15in}

\noindent {\em See also}\footnote%
{in Encyclopedia of Nonlinear Science, cited above.}
\textbf{Einstein equations; Hodograph transform; Matter, nonlinear theory of; String theory}

\def\refname{\normalsize Further Reading}

\end{document}